\documentstyle{l-aa}
\begin{document}
%
%
\thesaurus{ 11.03.1; 12.04.1; 12.12.1 }
%
%
\title{ Subcluster merging in clusters of galaxies
              and the cosmological density parameter }
\author{ F. E. Nakamura\inst{1}
     \and  M. Hattori\inst{2}
     \and  S. Mineshige\inst{1}
}
\offprints{ F. E. Nakamura }

\institute{ Department of Astronomy, Faculty of Science,
        Kyoto University, Sakyo-ku, Kyoto 606-01, Japan
   \and Max-Planck-Institut f\"ur Extraterrestrische Physik,
        D-85740 Garching, Germany
}
\date{ Received July 7,1994 , accepted February 22,1995 }

\maketitle

%
%
\begin{abstract}
Recent X-ray observations have established that
collisions between subclusters of galaxies are rather common phenomena.
Prompted by such observations,
we have performed N-body simulations of
two equal-mass subclusters of galaxies, which are going to merge.
We first have confirmed that
only a part of kinetic energy associated with
the relative motion of two subclusters
is converted to internal energy of each subcluster at the first encounter.
Two subclusters, therefore, once pass through each other,
turn around due to gravitational attraction,
and finally merge at the second or third encounter.
We performed simulations for a variety of dark-matter (DM) distributions,
and find that the time scale for washing out the double peak structures after
the
first encounter strongly depends on the distribution of the density and
velocity
distributions of the DM.
It takes longer when the DM is
spatially extended and/or when
the velocity distribution of the DM has a Gaussian
shape, rather than a uniform distribution.
According to our calculation
it takes more than $4 \times 10^9$yr  after the first encounter until the
density contour shows only a
single peak.

To explain the high fraction of the clusters with substructures among
nearby clusters, Richstone,Loeb \& Turner(1992; hereafter RLT)
required recent($\sim$
$10^9 h^{-1}$yr) cluster formation. However, our results show that
the timescale for subcluster merging is still uncertain
and possibly much longer
than the time scale assumed by RLT.
Caution should be exercised when concluding that the density of the universe is
high by using RLT's method.
\keywords{ Galaxies: clustering -- dark matter -- large-scale
structure of Universe }
\end{abstract}

\section{Introduction}
The existence of double or multiple peaks in the spatial mass distribution
has been established for a number of clusters of galaxies.
Jones \& Forman (1990), for example, have
reported that about 30 \% of 250 clusters of galaxies
have double or multiple peaks in the X-ray
surface brightness.
Since the X-ray emissivity is more sensitive to the density rather
than to the temperature,
their observations strongly indicate the presence of multiple peaks
in the distribution of the intracluster medium (ICM).
ROSAT observations of the clusters of galaxies have established that the high
fraction of the galaxy clusters have double or multiple peaks in the
distribution of
the X-ray emitting gas (e.g. Briel et al. 1991).
The existence of such double or multiple peaks
has been also known for the galaxy distribution
in $\sim$ 40 \% of 40 clusters of galaxies (Geller \& Beers 1982).
Moreover, Miralda-Escud\'{e} (1991) has recently inferred from the
analysis of the gravitational lensing clusters
that the distribution of the gravitational mass, that is the mass of the all
constituents including dark matter, in the clusters of galaxies is not
spherically
symmetric, but has multiple peaks. These observations of the distribution
of  ICM, galaxies and dark matter in clusters suggest frequent
occurrence of merger events between subclusters in
clusters of galaxies.

Frequent collisions between subclusters are of great importance for
measuring the geometry of the universe.  Regarding the presence of
substructuring as a signature of recent cluster formation, RLT estimated the
lower limit to the
cosmological density parameter to be $\Omega_0 \ga 0.5$.  This is a
rather large value and is in contradiction with the results deduced
from the another methods (e.g. Briel et al. 1991; Yoshii \& Takahara
1989).  For example, the local estimates of the cosmological density
parameter, the baryon fraction in individual objects (e.g. Briel
et al. 1992; White et al. 1993)
and the results from the peculiar velocity field (e.g.
Tully 1987), and deep galaxy number counts (Yoshii \& Takahara
1989) support a low density universe.   We thus need to
re-examine the basic assumptions made by RLT and check the significance
of their obtained result.

Their fundamental assumptions are:
(1) the initial perturbations are comoving with the Hubble flow,
(2) the shape of the initial perturbations is spherically symmetric,
and (3) the relaxation
timescale, $\tau_{\rm relax}$, on which multiple structures within a cluster is
erased after the gravitational collapse of the cluster, is at most a cluster
crossing time, 
of the order of $\sim 10^9 h^{-1}$ yr.
The first and second assumptions have been checked by Bartelmann et al. (1993).
They introduced initial velocity fluctuation and non-spherical
perturbation to find that RLT's conclusion gets firmer
by these additional and more realistic conditions.

However, the third assumption seems to be more crucial for RLT's
argument since the lower limit of the cosmological density parameter
obtained by RLT's method is rather sensitive to the assumed relaxation
time will be shown in
Sec.4.   McGlynn \& Fabian (1984, hereafter MF) have shown that
it takes more than several crossing times
after the first encounter of the substructures to erase
multiple structures after the first encounter of
the substructures unless a
strong two body relaxation works during the collision.

In the present study, we perform extensive studies of
subcluster merging, following the line of MF,
but adopting more general conditions for the initial spatial and
velocity distributions for the DM in clusters.  In fact,
the relaxation time depends strongly on these distributions and
the spatial distribution of the DM is still largely uncertain (see Sec.2.2).
Using the simulation results, we investigate how the constraint on
the cosmological density parameter using RLT's method is relaxed.
In Sect. 2 we briefly describe our model.
The results of N-body simulations are presented in Sect. 3.
The final section is devoted to summary and discussion.

\section{Models}
\subsection{Basic assumptions}
We investigate the merging of two virialized subclusters as an example
for substructure within an unrelaxed cluster of galaxies.
We use the same model parameters as MF
to enable a direct comparison (see Table 1).
Two subclusters with equal total masses are assumed,
since the size of substructure is comparable to that of the main structure
in about half of the observed clusters (Dressler \& Shectman 1988).
Particle number ($N$) of each cluster is 1024,
20 \% of which are assumed to be galaxies and the remaining 80 \% to DM.
The mass of a dark-matter particle, $m_{\rm DM}$, is 100 times larger than
the mass of a galaxy particle, $m_{\rm gal}$.

%
%
\begin{table}
\caption[ ]{
Standard model parameters.
$N$ represents the total number of galaxy or DM particles,
$M$ is the total mass of one subcluster,
and $m$ is a mass of one galaxy or DM particle.
Indices DM and gal represent
the dark matter and the galaxy component, respectively.
}
 \begin{flushleft}
  \begin{tabular}{|l|l|}
\noalign{\smallskip}
\hline
 Particles~($N$)          &  $ 1024 \times 2 $ \\
 Total mass~($M$)         &  $ (8 \times 10^{14}M_\odot) \times 2 $ \\
 $N_{\rm DM}/N_{\rm gal}$ &  4  \\
 $m_{\rm DM}/m_{\rm gal}$ &  100 \\
 $M_{\rm DM}/M_{\rm gal}$ &  400 \\
\noalign{\smallskip}
\hline
  \end{tabular}
 \end{flushleft}
\end{table}

Here we assume that the gravitational force of the ICM is negligible for
the evolution of the cluster.

\subsection{Initial mass distributions in each cluster}
It is observationally known that the galaxy distribution in compact clusters
is well approximated by the isothermal King distribution (King 1972),
\begin{equation}
  \rho_{\rm gal}
    = {\rho_{\rm gal}^0\over \Bigl[1+(r/r_{\rm gal})^2\Bigr]^{3/2}},
\end{equation}
where $\rho_{\rm gal}^0$
is the central density and $r_{\rm gal}$ is the core radius
for the galaxies in the cluster (Sarazin 1988).
According to Bahcall (1975), we fix the core radius to
$r_{\rm gal} = 250$ kpc in the present studies.
The central density
$\rho_{\rm gal}^0$ is defined by
$\int 4\pi \rho_{\rm gal}(r) r^2dr = M_{\rm gal}$,
where $M_{\rm gal}$ is the total mass of the galaxies
in the cluster (Table 1).
Since this integral diverges  $M_{rm gal} \sim \ln r$ at large $r$,
we truncate the distribution at $r_{\rm max} = 5 r_{\rm gal}$,
which is 1.25 Mpc for $r_{\rm gal} = 250$ kpc.

%
%
\def\rdg{$r_{\rm 0g}/r_{\rm 0D}$}
\def\alp{$\alpha$}
\def\afo{$ 4/3 $}
\def\ath{$  1  $}
\def\atw{$ 2/3 $}
\begin{table}
\caption[ ]{
Adopted parameters for the structure of a
subcluster and the resultant rebounce times (1 Gyr = $10^9$yr).
The first column is the model names.
$\rho(r)$ is the density distribution of the DM.
MK, IK and PL represent the modified King, the isothermal King
and the power-law density distributions, respectively.
$\alpha$ is the slope of the density profile.
$r_{\rm DM}/r_{rm gal}$ is the core radius of the DM distribution
in unit of the core radius of the galaxy distribution.
$f(v)$ is the velocity profile of the DM.
G and U represent the Gaussian distribution and uniform distribution,
respectively.
The initial relative velocity is given by Eq. (5), except for Model J,
in which $V_{\rm ini} = 0$.
The core radius for galaxies, $r_{\rm gal}$, is taken to be 250 kpc,
while $R_{\rm ini} = 3.125$ Mpc in all models.
}
 \begin{flushleft}
  \begin{tabular}{|c|l|c|r|c|r|}
\noalign{\smallskip}
\hline
Model & $\rho(r)$ & \alp & ${r_{\rm DM}/r_{\rm gal}}$ & $f(v)$
& $\tau_{\rm reb}$(Gyr) \\ \hline
 A& MK   & \atw&  $0.1$ & G & 9.0  \\
 B& MK   & \atw&  $1.0$ & G & 8.0  \\
 C& MK   & \atw&  $3.0$ & G & 9.5  \\
 D& IK   & \ath&  $0.1$ & G & 7.0  \\
 E& IK   & \ath&  $1.0$ & G & 8.5  \\
 F& IK   & \ath&  $3.0$ & G &10.5  \\
 G& MK   & \afo&  $0.1$ & G & 6.0  \\
 H& MK   & \afo&  $1.0$ & G & 7.5  \\
 I& MK   & \afo&  $3.0$ & G & 9.0  \\
 J& MK   & \afo&  $0.1$ & G & 4.0  \\
 M$_0$ & PL &(2/3) &(0.0) & U & 7.0 \\
 M$_1$ & PL &(2/3) &(0.0) & G &11.0 \\
\noalign{\smallskip}
\hline
  \end{tabular}
 \end{flushleft}
\end{table}

Unlike the galaxy distribution,
the DM distribution is still under discussion (Hughes
1989; Fitchett 1990). We hence adopt the following parameterization in analogy
with the gas distribution in the cluster
(Cavaliere \& Fusco-Femiano 1976);
\begin{equation}
    \rho_{\rm DM}
    = {\rho_{\rm DM}^0\over \Bigl[1+(r/r_{\rm DM})^2\Bigr]^{3\alpha/2}}
\end{equation}
where the core radius, $r_{\rm DM}$, and the slope, $\alpha$, of the DM
distribution are free parameters.
The adopted parameters are summarized in
Table 2.  A constant $\rho_{\rm DM}^0$ is determined so that $M_{\rm DM} = \int
4\pi \rho_{\rm DM}(r) r^2dr$. The combined X-ray and optical study of the mass
distribution in the Coma cluster (Hughes 1989) has shown that the
mass-follows-light models are the most precise to date, i.e. $r_{\rm DM}=r_{\rm
gal}$ and $\alpha =1$ in our notation. The study of the distribution of the
dark matter in the A2256 (Henry et al. 1993) has shown, on the other
hand, that the mass-follows-gas models, i.e. flatter distribution with
$\alpha\sim 2/3$, is also consistent with the present
X-ray observations.
A steeper mass distribution than the galaxy distribution, i.e. $\alpha\sim
4/3$, has not been rejected either (Hughes 1989; Ponman 1992).
Further more, much smaller core radii for the DM distribution in
rich cluster than the core radii of the luminous components
are inferred from the mass distribution measurements using the
gravitational lensing techniques;
the high rate of finding arc-like structures (Le Favre et al. 1994; Wu \&
Hammer
1993) and the geometrical study of the giant luminous arcs in distant clusters
(e.g. Hammer 1991; Miralda-Escud\'{e} \& Babul 1994; Wu 1994).
For comparison, we also performed a simulation with a
 large core radius, i.e.
$3\times r_{\rm gal}$.  Note that MF adopted a power-law density profile
\begin{equation}
  \rho_{\rm DM} = \rho_{\rm DM}^0 \Bigl(r/r_{\rm gal}\Bigr)^{-2},
\end{equation}
corresponding to a profile with $r_{\rm DM} = 0$ and $\alpha = 2/3$ in our
notation.

We assume two different velocity distributions:
One is a Gaussian distribution
with a cut-off at the velocity of $V = \pm 4\sigma$.
$\sigma$ is determined by the condition
that the initial system is virialized.  The second one is a uniform velocity
distribution,
from $V = 0$ to $V = V_{\rm max}$, where $V_{\rm max}$ is determined
by the condition that the initial system is virialized.
The latter one was used by MF.
The adopted parameters of all the calculated models are summarized in Table 2.

\subsection{Time evolutionary calculations of subclusters}
In a flat universe, the maximum expansion radius of perturbations which
collapse at a redshift $z_c$ in a $\Omega_0=1$ universe is expressed as
\begin{equation}
  \ell_{\rm max} = 6.0 M_{15}^{1/3} (1+z_c)^{-1} h^{-2/3}{\rm Mpc},
\end{equation}
where $M_{15}$ is the mass of the perturbation in units of
$10^{15}$M$_\odot$.
For $z_c \sim 1.0$ ($\sim 4.0 \times 10^9h^{-1}$ yr from now in a
critical density universe and even more in a low density universes), we
find $\ell_{\rm max} \sim 3.0 h^{-2/3}$ Mpc. If $\Omega_0<1$ or $h<1$, the
maximum expansion radius of the clusters which collapsed at $z_c = 1$ is larger
than the above value.  The initial separation of two clusters
is set to be $R_{\rm ini} = 2.5 \times r_{\rm max} (=3.125$ Mpc) in all our
models.
Zero initial relative velocity with respect to the center of mass of
the system , $V_{\rm ini}=0$ (model J), corresponds to a collapse
 earlier than $z_c = 1$.
Larger $V_{\rm ini}$ correspond to the case of the larger
maximum expansion radius, i.e. a later collapse.

In all our calculations, except for model J,
\begin{eqnarray}
 V_{\rm ini} & =      & \sqrt{2GM\over R} \nonumber \\
             & \simeq &
                        1700 ({\rm km/s})
                        \Bigl({M\over 8\times 10^{14}{\rm M_\odot}}\Bigr)^{1/2}
                        \Bigl({R_{\rm ini} \over 3 {\rm Mpc}}\Bigr)^{-1/2}.
\end{eqnarray}
This is derived from the condition that two clusters were at rest at infinity.
The crossing timescale, except for model J, is then approximately
\begin{eqnarray}
 t_{\rm c} & \equiv & R_{\rm ini}/ V_{\rm ini} \nonumber \\
           &   =    & 1.0 \times 10^9
               \Bigl({M\over 8\times 10^{14}{\rm M_\odot}}\Bigr)^{-1/2}
               \Bigl({R_{\rm ini} \over 3 {\rm Mpc}}\Bigr)^{3/2} {\rm yr}.
\end{eqnarray}

The gravitational force is calculated by direct summation.
We use softening parameter of $\epsilon = 0.6 $ Mpc.
It has been shown (see p36-38 in Sarazin 1988) that
the dark matter cannot be bound
to individual galaxies but rather must form a continuum to prevent clusters
from over relaxation due to dynamical friction.
In other words, the two-body relaxation timescale in clusters is much longer
than the timescale of a merger.
Therefore the merging of clusters is treated as a
collisionless process and we took a large value of
$\epsilon=0.6$Mpc which saves a lot of computating time.

The Equations of motion of particles are
integrated by the 4-th order Runge-Kutta method.
To examine the stability of the initial cluster configurations,
we calculate the evolution of a single cluster for several crossing
times (Eq. (6)), confirming no appreciable structural changes.
The energy is conserved within 3\%
of the total energy in the every calculation.

\section{Simulations}
\subsection{General behavior}

We first calculate models with $\alpha = 2/3$,
the same density profile as those of MF,
$\rho_{\rm DM} \propto r^{-2}$ at $r \gg r_{\rm gal}$.
The core radius is assumed to be $r_{\rm DM}/r_{\rm gal} =$ 0.1, 1.0, and 3.0,
for models A, B, and C, respectively (see Table 2),
and the initial velocity is given by Eq. (5).
Figure 1\footnote{The number of particles in Fig. 1 is
reduced for the sake of saving the size of PS file in
this online preprint.} shows a three-dimensional distribution of the
DM particles for model B.
Two clusters pass through each other,
turn around due to mutual attraction,
and recollide.

\begin{figure}
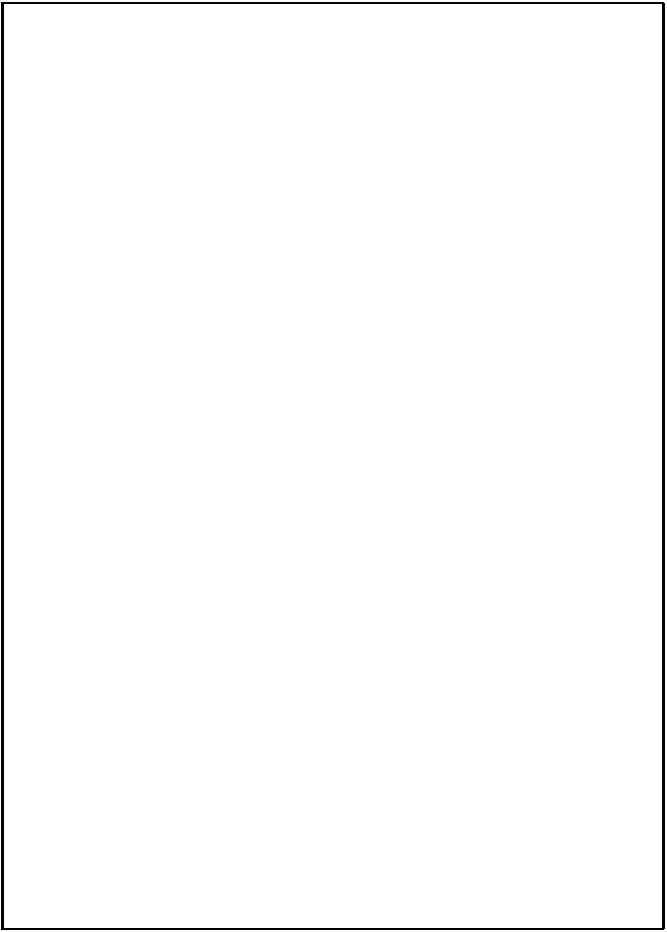

\picplace{12.3 cm}
\caption[ ]{
Three-dimensional distribution of the
DM particle positions for model B.
The size of the box is 3 Mpc $\times$ 3 Mpc $\times$ 8 Mpc.
The elapsed times in unit of $t_{rm cr}$ are, from top to bottom,
$t = 0.0, 1.6, 3.2,$ and 4.8 in the left panel, and
$t = 6.4, 8.0, 9.6$, and 11.2 in the right panel.
}
\end{figure}

The upper panel of Fig. 2
illustrates the time variations of the distance of the
center of mass of the two clusters in units of
the crossing time, $t_{\rm cr}$ (Eq. (6)).
The lower panel of Fig. 2 displays
the half-mass radius of the DM in a cluster evolving with time,
which is a rough indicator of the kinetic energy associated with
the internal motion.
The half-mass radius increases right after the encounter.
This indicates that the energy associated with
the relative motion of the clusters
is partly transferred into internal energy within the clusters
during the first
encounter.
Nevertheless, further encounters are necessary to damp
the oscillating motions of two subclusters completely,
by forming a big cluster.

\begin{figure}
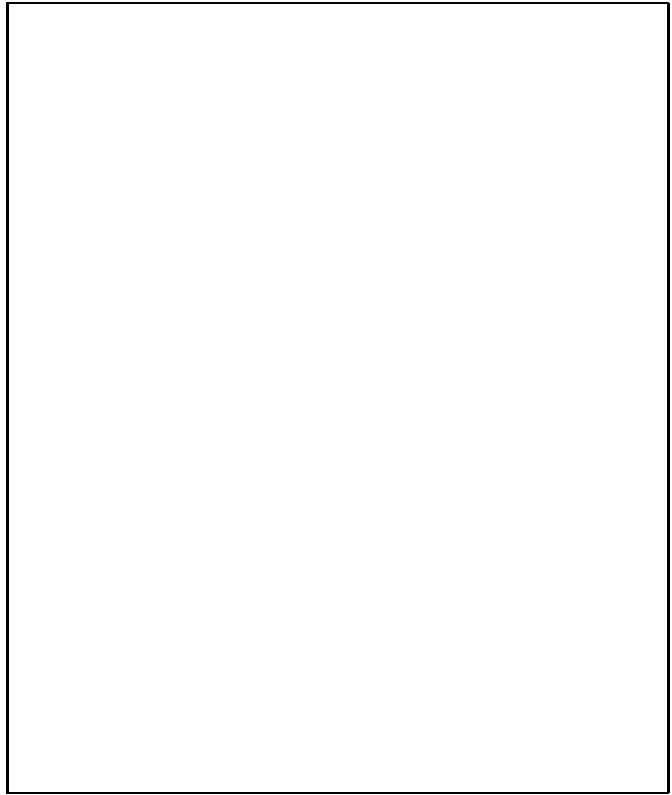

\picplace{10.5 cm}
\caption[ ]{
The relative distance between the centers of the DM distributions
of the two subclusters (the upper panel) and in
the half-mass radius of the DM in one cluster (the lower panel)
for models with $\alpha = 2/3$ versus time.
In both panels, the solid line corresponds to the model with
$r_{\rm DM}=3 r_{\rm gal}$, the dotted line to $r_{\rm DM}=r_{\rm gal}$,
and the dashed line to $r_{\rm DM}=0.1r_{\rm gal}$, respectively.
The units of time and distance are $t_{\rm c}$
(Eq. (4)) and 1 Mpc, respectively.
}
\end{figure}

Our main aim of this paper is to investigate the question how long the bimodal
structure is preserved
after the gravitational collapse of the cluster.
The time of the first encounter of two substructures in our simulations
corresponds to the cluster collapse.
Since at least two encounters are necessary to wash out
the bimodal structure in
our all simulated models,
the time interval between the first and second encounter,
which we define as the rebounce time,
gives a lower limit to the merging time-scale
after the gravitational collapse.
The resultant rebounce times are summarized in Table 2.
We see that it is insensitive to the core radius of the DM
in the models with $\alpha = 2/3$.

\subsection{Comparison with MF}

For comparison with the results of MF, we performed
similar simulations in model M$_0$.
The density profile is power-law (Eq. (3))
(corresponding to the models with $\alpha = 2/3$ and $r_{\rm DM}=0$) and
the velocity distribution of the DM is uniform.
Figure 3 displays the distance of the two subclusters
and the half-mass radius, respectively,
as a function of time.
The rebounce time of model M$_0$ is $7\times 10^9$yr.
Since the time unit defined in MF corresponds to $3\times 10^9$yr in our
models,
this rebounce time corresponds to 2.3 in MF's time units.
This is similar to the rebounce time of model A in MF which is the
most similar to our models.
Therefore, our simulation code could well reproduce MF's results.

\begin{figure}
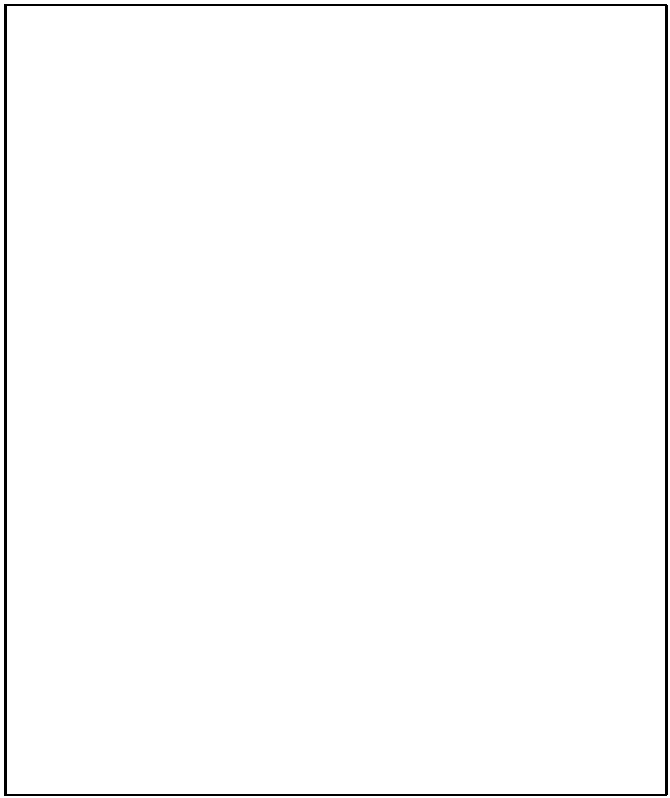

\picplace{10.5cm}
\caption[ ]{
The same as Fig. 2 but for the models without cores.
The dotted line corresponds to MF's original model with a flat
velocity profile (model M$_0$),
while the solid line represents the model with the Gaussian velocity
distribution (model M$_1$)
}
\end{figure}

\begin{figure}
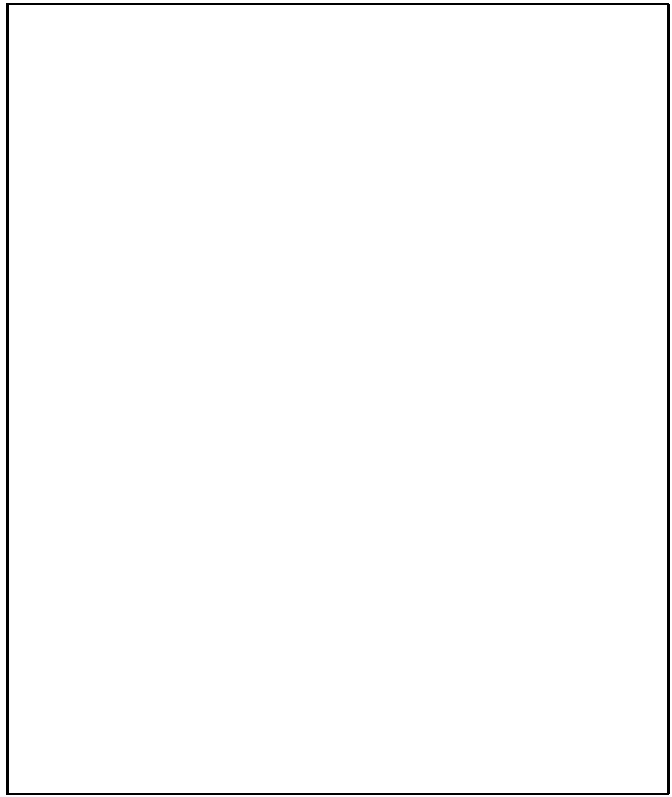

\picplace{10.5 cm}
\caption[ ]{
The same as Fig. 2 but for $\alpha = 1$.
}
\end{figure}

\begin{figure}
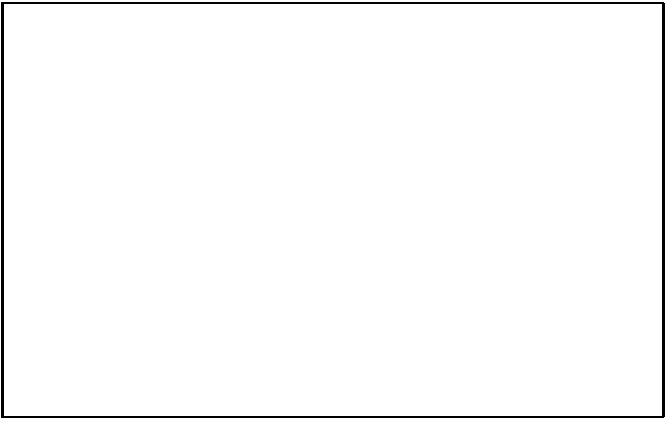

\picplace{5.5 cm}
\caption[ ]{
The same as upper panel of Fig. 2, but for $\alpha = 4/3$
and $r_{\rm DM} / r_{\rm gal} = 0.1$.
The dashed and solid lines represent
the case with non-zero and zero initial velocity, respectively.
The unit of time is $10^9$yr.
}
\end{figure}

To investigate the effect of
velocity distribution of the DM,
a more realistic velocity distribution, a Gaussian distribution,
is used in M$_1$,
that is the Gaussian distribution.
The rebounce time is about 1.5 times longer
than MF's model M$_0$.

\subsection{Models with $\alpha = 1$ and 4/3}
In models with steeper density profiles,
the rebounce time is rather
sensitive to the core radius.
Figure 4 is the same as Fig. 2 but for models with $\alpha = 1$;
the solid line represents a model with $ r_{\rm DM} = 3 r_{\rm gal}$,
the dashed line a model with $ r_{\rm DM} = 0.1 r_{\rm gal}$.
The rebounce time responds with the core radius much more than the case with
$\alpha=2/3$; $t_{\rm reb}$ is reduced by a factor of 1.5, when
we decrease $r_{\rm DM}$ by one order of magnitude.
A similar trend can be seen for the case of $\alpha = 4/3$ (see Table 2).
The core-radius dependence becomes weaker as $\alpha$ increases.

\subsection{Bound models: dependence on $V_{\rm ini}$}

As discussed in Sect. 2.3, the case of the zero initial relative velocity
correspond to the case in which the cluster collapses at earlier than $z_c =
1$. To demonstrate that multiple-peak structures  still persist in the present
mass distribution 
even if clusters collapsed at such an early epoch,
we calculate a model with $V_{\rm ini}=0$.
Since the rebounce time of model G is the shortest among our calculated models,
we set the same initial conditions as model G except $V_{\rm ini}$ in model J.
The rebounce time for models
with $V_{\rm ini}=0$ but with other combinations of $\alpha$
and $r_{\rm DM}$ should be longer than the result of model J.

\begin{figure}
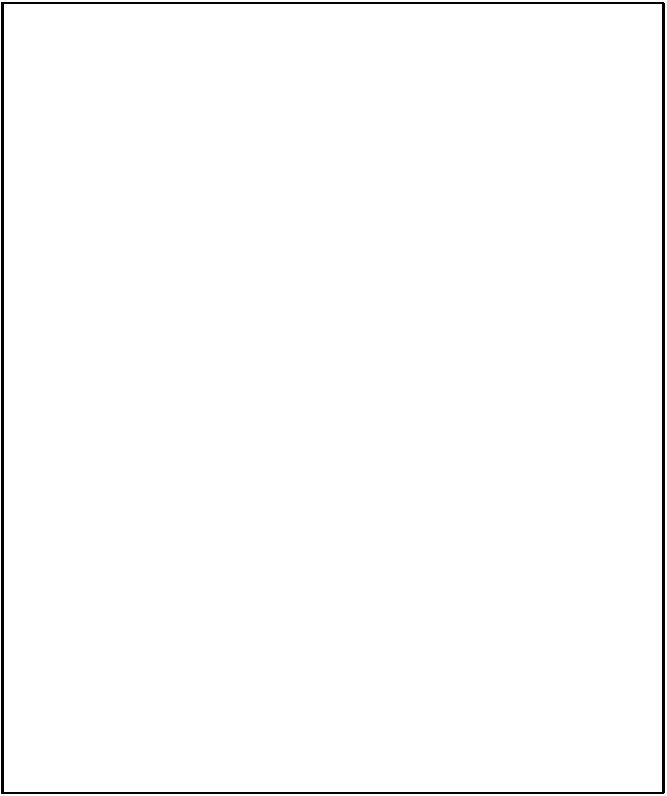

\picplace{10.5 cm}
\caption[ ]{
Density distribution of some calculated models.
The upper panels shows  models with $\alpha = 2/3$ together with model
M0, while the lower panels shows models with $\alpha = 4/3$.
In both panels, the solid line corresponds to
the case with
$r_{\rm DM}=3 r_{\rm gal}$, the dotted line $r_{\rm DM}=r_{\rm gal}$, the
dashed line
$r_{\rm DM}=0.1r_{\rm gal}$, and the dash-dotted line M0, respectively.
}
\end{figure}

The time variation of the separation is illustrated in Fig. 5.
Clearly, the rebounce time decreases by a $\sim 30$ \%, however, still it is
factor $4h$ longer than the relaxation time assumed by RLT,
that is $10^9h^{-1}$yr.
Further oscillatory motion with an amplitude of 1Mpc still remains
after the second encounter.
This result shows that we do not have to require
the recent formation of nearby clusters with substructures
as RLT did, where they
assumed the substructuring clusters have collapsed
within $10^9h^{-1}$yr from now.

\subsection{Interpretation of the results}

Here we try to give the physical interpretations of the above results.
As noted in Sect. 2.3, the two-body relaxation is not effective for subcluster
mergers.
Instead, violent relaxation,
an energy exchange through variations of the potentials,
plays a dominant role.
The efficiency of the energy conversion
then critically depends on how the potential changes with time,
which, in turn, depends on the precise DM distribution in the cluster.
The specific energy of a particle, $e = {v^2}/2 + \phi$, varies as
\begin{equation}
 {{de}\over{dt}}=v{{dv}\over{dt}}+ {{\partial \phi} \over {\partial t}}
+ v \nabla \phi
 = {{\partial \phi} \over {\partial t}},
\end{equation}
where $v$ is the velocity of a particle,
$\phi$ is the gravitational potential,
and we used equation of motion, $dv/dt = -\nabla \phi$ (Lynden-Bell 1967).
During the subcluster collisions, moreover, we may approximate as
$\partial \phi / \partial t \simeq (\partial\phi/\partial x) v$.
That is, the steeper is the potential curve, the quicker proceeds
the merging process due to an efficient energy conversion.

In Fig. 6
we illustrate the density profiles of some calculated models,
whereas Fig. 7 displays a potential of each model cluster.
For the same core radius, say, $r_{\rm DM}/r_{\rm gal} = 0.1$,
the potential curve is steeper, when $\alpha$ is larger,
so that the relaxation time becomes shorter
for models with larger $\alpha$ values.
When a cluster has a relatively large core radius, $r_{\rm DM}$,
however, differences in potential curves are not significant,
which explains why $t_{\rm reb}$ does not change in models
with $r_{\rm DM}/r_{\rm gal} = 1.0$ or 3.0.

\begin{figure}
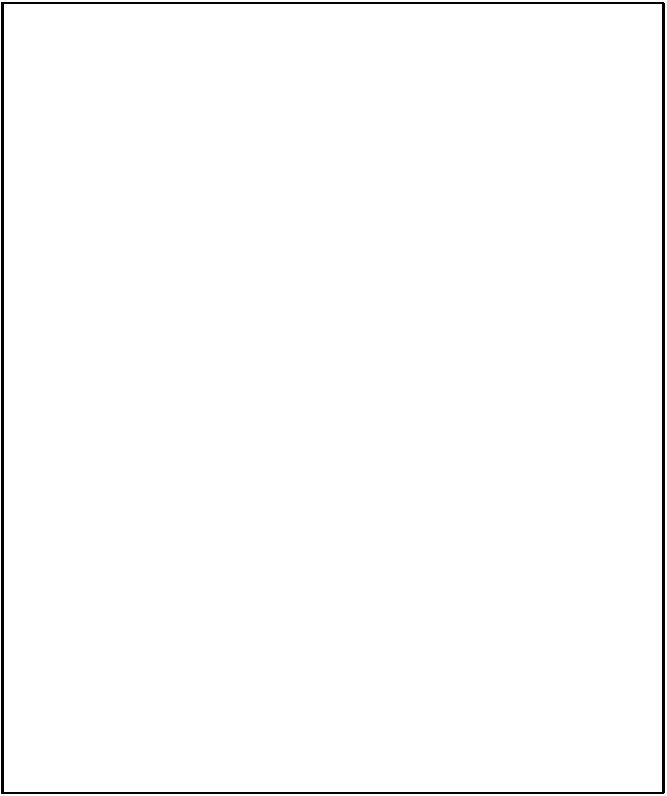

\picplace{10.5 cm}
\caption[ ]{
Gravitational potentials generated by one cluster
for each mass distribution model. Lines are
same as Fig. 2.
}
\end{figure}

Likewise, for the same $\alpha$ values, say, $\alpha = 4/3$,
the potential curve is steeper, when $r_{\rm DM}$ is smaller.
The relaxation time is thus a decreasing function of the core radius
when $\alpha = 1$ or 4/3.
In the models with $\alpha=2/3$, on the other hand,
this tendency is not clear, because the potential gradients
do not differ so much from each other.

We also anticipate that the relaxation time is shorter,
when the fraction of high-velocity particles is large.
Funato et al. (1992) have shown that particles with initially higher energy get
more energy than initially lower energy particles during the so called violent
relaxation process.
In this sense, a flat velocity distribution is favoured
than the Gaussian distribution in terms of triggering a rapid relaxation. The
distinction between models M$_0$ and M$_1$ is thus explained.

\subsection{Non-central collisions}

So far only head-on collisions of subclusters were considered.
We have also performed the simulations for an impact parameter
$b = 250$ kpc. We find an even longer rebounce
time than in central collision.
This can be explained as follows:
For $b \neq 0$, the center of each cluster
does not pass through the center of the other component
where the potential is the deepest.
This means, each particle experiences a smaller potential variation
than the case with $b = 0$.
Since the energy-conversion efficiency from the bulk motion
to the internal motion depends critically on the potential changes
(Eq. (7)), the relaxation times are systematically longer when
$b \neq 0$, compared with the models with $b = 0$.
Therefore, a head-on collision yields a shorter
relaxation time than an offcenter collision.
The possibility of non-zero
impact parameter in real cluster collision (Fabian \& Daines 1991)
strengthens our
conclusion mentioned in Sec.3.4.

\section{Discussion and conclusion}

We have performed N-body simulations to study the relaxation process of
two, equal-mass colliding subcluster of galaxies.
We have demonstrated that it is difficult
to erase the bimodal structures in the dark matter distributions at the
first encounter and that it remains for a long period after
the first encounter.
Our results
demonstrate that the rebounce time (hence the relaxation time) depends
on the precise density and velocity distribution of the DM.  There are two
key parameters for the density profile, the core radius and the power
index of the DM distribution in the outer region.  We find that a
smaller core radius or a steeper DM
distribution yields a quicker relaxation.  This is
understood as the result of violent relaxation, in which a more efficient
energy conversion from the relative motion to the internal motion occurs when
the potential is steeper; namely, when a core radius is smaller and
the density distribution is steeper.
Our results show that it takes more than $4h$ times longer
than the time scale assumed by RLT to erase substructures in the clusters after
the first encounter even in the most quick merging cases, if the cluster
collapsed later than a redshift of 1.

\begin{figure}
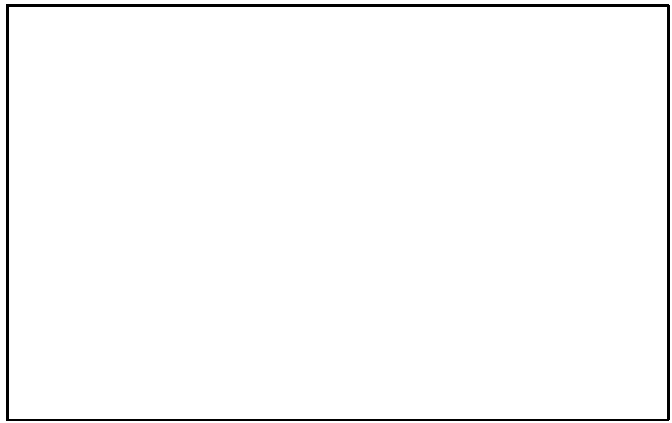

\picplace{5.5 cm}
\caption[ ]{
Relation between $\Omega_0$ and $\delta t$
for various values of $\delta \tilde{F}$,
the fraction of clusters formed between $T_0 - \delta t$ and
$T_0$ with $T_0$ being the present age of the universe (cf. RLT).
}
\end{figure}

This large merging time scale
results in an appreciable decrease in the lower limit of $\Omega_0$.
Figure 8 displays the relation between $\Omega_0$ and $\delta t$
for various values of $\delta \tilde{F}$,
which represents the fraction of clusters
formed between $T_0 - \delta t$ and $T_0$,
where $T_0$ is the present age of the Universe (cf. RLT).
We find an approximate fitting formula from Fig. 8
\begin{equation}
 \Omega_0 \ga {0.32 \delta \tilde{F} \over {\delta t/T_0(\Omega_0,\lambda_0)}}
\end{equation}
for 0.25 $\la \delta \tilde{F} \la 0.40 $ and
$ 0.0 \leq \delta t / T_0 \la 0.6 $.
Numerically solving Eq. (8) for $\Omega_0$ yields an
approximate formula,
\begin{equation}
     \Omega_0 \ga {0.26 \delta \tilde{F} \over \delta t~H_0},
\end{equation}
which is almost independent on $\lambda_0$.
It shows that
the derived lower limit of $\Omega_0$ is strongly depending on the
assumed relaxation time.
RLT took $\delta \tilde{F} \ga 0.25$ mainly from observed
X-ray surface brightness contours by Jones \& Forman(1990)
and derived $\Omega_0 \ga 0.5$ for $\delta t \la 0.1 H_0^{-1}$.
Our simulations
demonstrate that the rebounce timescale
is $t_{\rm reb}/t_{\rm RLT}\sim 4-10h$, where $t_{\rm
RLT}=10^9h^{-1}$yr is merging timescale adopted by RLT.
Since the relaxation time is longer than the rebounce
timescale, we have
\begin{equation}
  \delta t \ga t_{\rm reb} \simeq (4 - 10) \times
10^9{\rm yr},
\end{equation}
as the relaxation time of the DM distribution.
If the gas always follows the DM distribution, we can apply this result to
estimate the lower limit of $\Omega_0$ with this substructure frequency
deduced from X-ray observations.
It shows that even if
$\delta t=t_{\rm reb}$, the derived lower limit of the density of the universe
is more than a factor of 2 smaller
than the value derived by RLT as
\begin{equation}
    \Omega_0 \ga 0.18
            \Bigl({t_{\rm reb}/t_{\rm RLT}\over 4.0}\Bigr)^{-1}.
\end{equation}

If we take into account the fact that the oscillatory
motion of merging substructures still survives after the second
encounter and that the adopted value is obtained by using the
model with an extremely centrally concentrated mass distribution which leads
to the shortest relaxation time,
the true relaxation time could be longer than this  and
a much lower density universe could be consistent with the observed
substructure frequency.

Finally, we would like to note the applicability of our present results to the
estimation of the mean density of the universe based on the statistics of the
appearance of the substructures in clusters.
In the above discussion, we have assumed that the morphology of the X-ray
emitting gas always follows the morphology of the dark matter distributions.
However, it is still unclear whether the hot gas follows the DM or not.
There is
evidence which supports this assumption both in simulations
(Roettiger et al. 1993)
and in the real world (Briel et al. 1992).
Roettiger et al. (1993) have shown that
even after the encounter of two clusters the X-ray emitting gas almost traces
the
dark matter distributions and they have suggested that the substructure in the
Coma
cluster found by X-ray morphology (Briel et al. 1992) is in this post encounter
stage.  However, there is also a counter evidence against this assumption in
simulations (Evrard 1990; Schindler \& M\"uller 1993)
where the bimodal structure of the X-ray emitting gas disappears when
two substructures meet although the dark matter of the two substructures passes
through
each other (Evrard 1990; Schindler \& B\"ohringer 1993).  Probably the both
cases are
occurring in real clusters depending on their merging conditions but it is not
sure which is the dominant process.  If the fraction of
merger events represented by the former case is dominant, we can
safely assume the above mentioned assumption.  However, if the
fraction of the merging events represented by the latter case is
dominant, the frequency of substructures in dark matter distribution
deduced from X-ray data is highly underestimated.  Fortunately, there is now
a new useful tool to study the dark matter distribution in
clusters directly, that is gravitational lensing technique
(e.g. Fort \& Mellier 1994).  If the latter is the case, we will find a large
discrepancy of substructure frequency between the results deduced by
the gravitational lensing technique and the X-ray data, that is the
former will be much higher.  This will be a good test which of
above two processes is dominant process in real world.

The special nature of the initial conditions in our simulations would also
limit the
applicability of our results to the real cluster merging.
There have been extensive N-body calculations based on somewhat more
cosmologically
motivated initial conditions than are the present calculations (e.g. Evrard et
al.
1993; Crone, Evrard \& Richstone 1994;Jing et al. 1994).
Although their approaches are basically more
straightforward to compare their results with real cluster observations, their
present results are still largely limited by their numerical resolution, size
of the
simulation boxes, number of the simulated samples, baryon fraction, and so on.
Actually,
the results obtained by this approach are still controversial.
Jing et al. (1994) have obtained much a larger fraction of clusters having
substructures in a low density universe than Evrard et al. (1993) who found
that the
morphology of clusters in low-density universes are much more regular,
spherically
symmetric than those in the critical density universe.
In this situation, our approach has a complementary meaning to examine the
cluster
merging process.
Our present study is enough to make a caution to require recent cluster
formation to
explain the high fraction of nearby clusters having substructures and to
conclude
that the density of the universe prefers to be high by using RLT's method.

In conclusion, the result of a high density universe RLT derived
should be still in question because multiple X-ray clusters
cannot be easily considered as evidence for recent cluster formation
according to our simulation results.
Further studies to clarify the dependence on the relaxation
timescale is required.

\begin{acknowledgements}
The authors would like to thank Sabine Schindler for valuable
comments and critical reading.
All the simulations were carried out on the CRAY Y-MP2E/264
whose computation time was provided by the Supercomputer
Laboratory, Institute for Chemical Research, Kyoto University.
F.E.N. thanks the stuff of the department of Ibaraki University
for their hearty encouragement.
S.M. is grateful to Grants-in-Aid for the Scientific Research
of the Ministry of Education, Science and Culture of Japan
(No. 05242213, 05836017), and the Yamada Science Foundation.
M.H. has been supported by the special Researchers' Basic Science Program of
the Institute
of Physical and Chemical Research (RIKEN) and a post-doctoral fellow ship at
MPE.
\end{acknowledgements}

\end{document}